\documentclass[11pt,twoside]{article}
\usepackage{asp2010}
\usepackage{subfigure}

\resetcounters

\begin{document}

\title{Wolf-Rayet Central Stars of Planetary Nebulae:  Their Evolution and Properties}
\author{K. DePew$^1$, D.J. Frew$^{1}$, Q.A. Parker$^{1,2}$ and O. De Marco$^1$}
\affil{$^1$Department of Physics \& Astronomy, Macquarie University, Sydney, NSW 2109, Australia}
\affil{$^2$Australian Astronomical Observatory, Epping, NSW 1710, Australia}
%\affil{$^3$Institution Full Address for Author3}}

\begin{abstract}
Over the past decade, the number of planetary nebula central stars (CSPN) known to exhibit the Wolf-Rayet (WR) phenomenon has grown substantially.  Many of these discoveries have resulted from the Macquarie/AAO/Strasbourg H$\alpha$ (MASH) PN Survey.  While WR CSPN constitute a relatively rare stellar type ($\lesssim$10\% of CS), there are indications that the proportion of PN harbouring them may increase as spectroscopy of more central stars is carried out.  In addition, with new and better distances from the H$\alpha$ surface brightness-radius relationship of Frew (2008), we can attempt a dynamical age sequence which may provide insight into the evolution of these stars.  
\end{abstract}

\section{Introduction}

Among central stars of planetary nebulae (CSPN), there exists a class of H-deficient objects that exhibit high mass-loss rates ($\gtrsim$ 10$^{-6}$ M$_{\odot}$ yr$^{-1}$) due to strong, fast stellar winds.  Their spectra resemble those of massive Wolf-Rayet stars, but these CSPN evolve from low- and intermediate-mass main sequence stars instead of massive O stars.  They are designated as [WR] stars to differentiate them \citep{1981SSRv...28..227V}.   

The [WR] class is subdivided into an oxygen sequence (designated [WO]), a carbon sequence ([WC]) and two controversial subclasses, the [WN] and [WN/WC] types.  Nitrogen lines are enhanced in the latter two, accompanied in the [WN/WC]s by higher carbon abundances \citep{2010A&A...515A..83T}.  Massive WOs and WCs possess surface abundances that differ from those of WNs.  These stars are believed to represent an evolutionary sequence in which progressively deeper layers of the star are exposed through strong wind-driven mass loss \citep[e.g.][and references therein]{2007ARA&A..45..177C}.  Thus, WCs and WOs possess different surface abundances than WNs.  C/He ratios were at one time thought to be less in late-type [WC]s than in early types \citep[e.g.][]{2001Ap&SS.275...53D}, but as noted by \citet{2007ARA&A..45..177C}, the lack of common diagnostics between late and early types could be a source of error.  More recent studies find that chemical abundances appear to be similar across [WC]s and [WO]s \citep[He:C:O $\sim$ 50:40:10 by mass for both;][]{2008ASPC..391...83C}, and their subclasses are primarily distinguished by temperature and degree of ionization.  PB 8, the lone published [WN/WC] type, has a much higher mass fraction of H (40\%) than [WC]s and [WO]s \citep{2010A&A...515A..83T}.   

The first scenario proposed to explain the formation of [WR] CSPN was the so-called ``Born-Again Scenario'' \citep[e.g.][]{1979A&A....79..108S,1983ApJ...264..605I}, in which a CSPN undergoes a thermal pulse that throws it back into the AGB.  Possible [WR] evolutionary paths are differentiated according to the stage at which the thermal pulse occurs:  the Asymptotic Giant Branch Final Thermal Pulse (AFTP), which occurs at the end of the AGB phase; a Late Thermal Pulse (LTP), which occurs when the star has left the AGB but has not yet ceased H-burning; or a Very Late Thermal Pulse (VLTP), which occurs when the star is already on the WD cooling track. See \citet{2001Ap&SS.275...15H} for a more detailed discussion.  Formation through a binary interaction has also been proposed \citep[e.g.][]{2002PASP..114..602D,2008ASPC..391..209D}.  \citet{2010MNRAS.406..626H} conclude that emission-line stars are less likely than ``normal'' H-rich PN to be found in binary systems, or else have larger orbital separations.  This may suggest that [WR]s result from a binary merger.

A by-product of many searches for [WR]s among the CSPN are weak emission-line stars or WELS, some of which are also H-deficient \citep{2003AJ....126..887M}.  These objects exhibit weaker emission lines at many of the same high-ionization wavelengths as [WR]s, but are likely less massive, evolutionarily unrelated objects, as evidenced by differences in Galactic scale height $z$ \citep{depewwels}.  Overall, a near-complete volume-limited 1 kpc sample suggests that 7$\pm$3\% of CSPN belong to the [WR] class, and $\sim$20\% of CSPN are H-deficient \citep{frewpark10}. 

\section{New [WR]/WELS Discoveries}

Recently 33 new [WR]/WELS and emission-line star candidates have been discovered in the course of the Macquarie/AAO/Strasbourg H$\alpha$ \citep[MASH;][]{2006MNRAS.373...79P,2008MNRAS.384..525M} survey \citep{depew}.  Of these, 19 are objects first found by us within the MASH sample, and 14 are previously known objects whose [WR] or WELS nature was discovered serendipitously by us in the course of spectroscopic follow-up of MASH objects.  See \citet{depew} for observational details and preliminary analysis.  The 17 MASH [WR]s detailed in that paper, added to the 7 previously found in the MASH survey \citep{2001MNRAS.322..877M,2003MNRAS.341..961P,2003MNRAS.346..719M}, make a contribution of 24 new objects to the list of [WR]s by our group.  With the 6 serendipitous discoveries, this represents an increase of $\sim$40\% in known [WR]s.  Within the MASH sample, there is also one [WN] subclass object \citep[PM5;][]{2003MNRAS.346..719M} and one probable [WN/WC] object \citep[Abell 48;][]{a48}. 

%\begin{figure}[htp]
%\setlength{\intextsep}{1.0cm}
%%\begin{minipage}[t]{13cm}
%\begin{center}
%\subfigure[MPA1611-4356, one of the new MASH [WR] PN, classifiable as [WC4] or [WO4].  See \citet{depew}.]{\label{mpa1611-4356}}\includegraphics[clip=true,scale=0.1,angle=270]{mpa1611-4356.eps}
%\subfigure[Preliminary subclass-dynamical age relationship.  The subclass index corresponds to: [WO1]=1, \ldots, [WO4]=4, [WC4]=5, \ldots, [WC11]=12.]{\label{DynAge}}\includegraphics[clip=true,scale=0.2,angle=270]{dynagecrop.eps}
%\end{center}
%\caption{My figures.}
%\label{Stuff}
%%\end{minipage}
%\end{figure}

\begin{figure}
%\centering
\setlength{\intextsep}{1.0cm}
\begin{flushleft}
\mbox{\subfigure{\includegraphics[scale=0.242,angle=270]{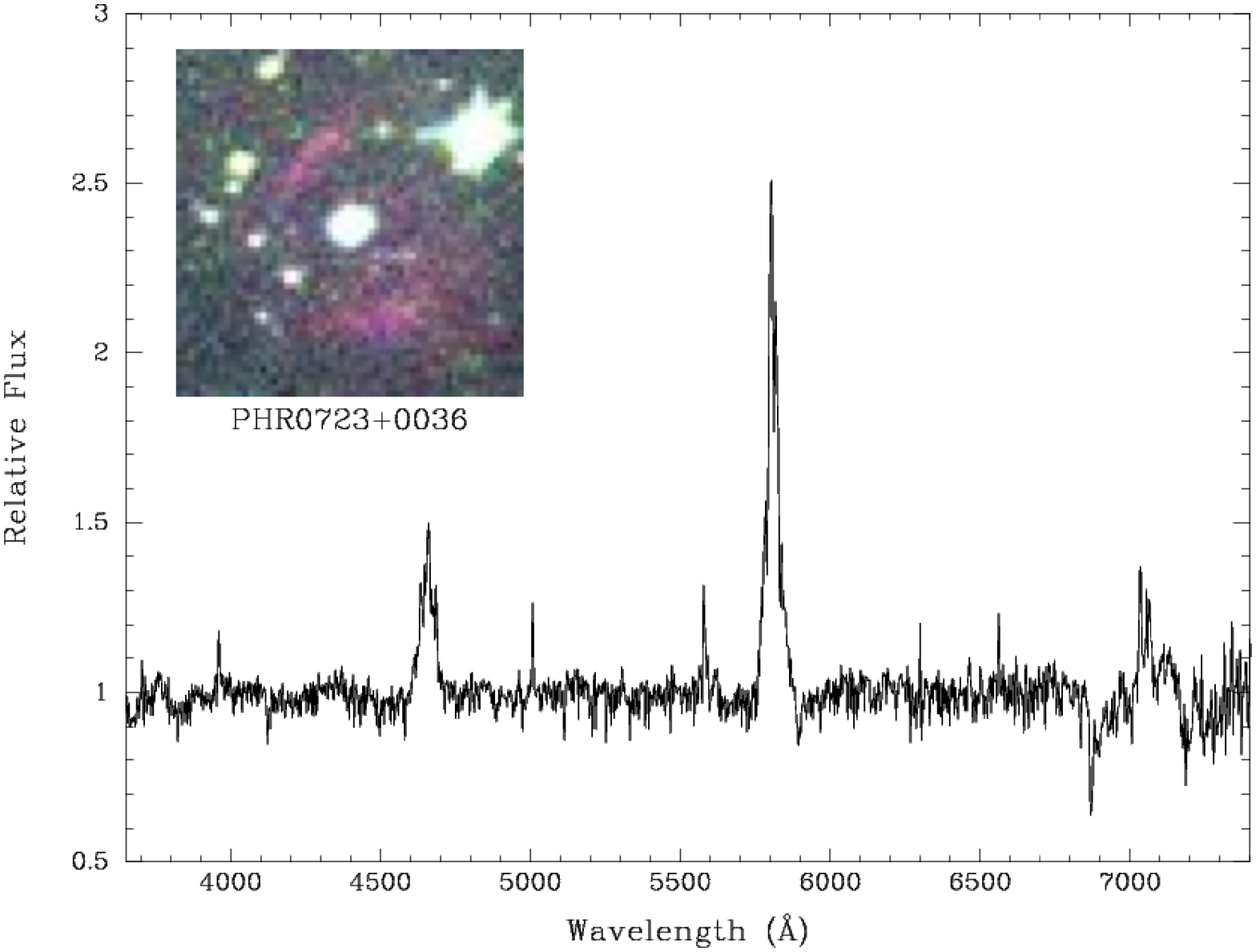}}\quad%scale=0.3,0.31
\subfigure{\includegraphics[scale=0.252,angle=270]{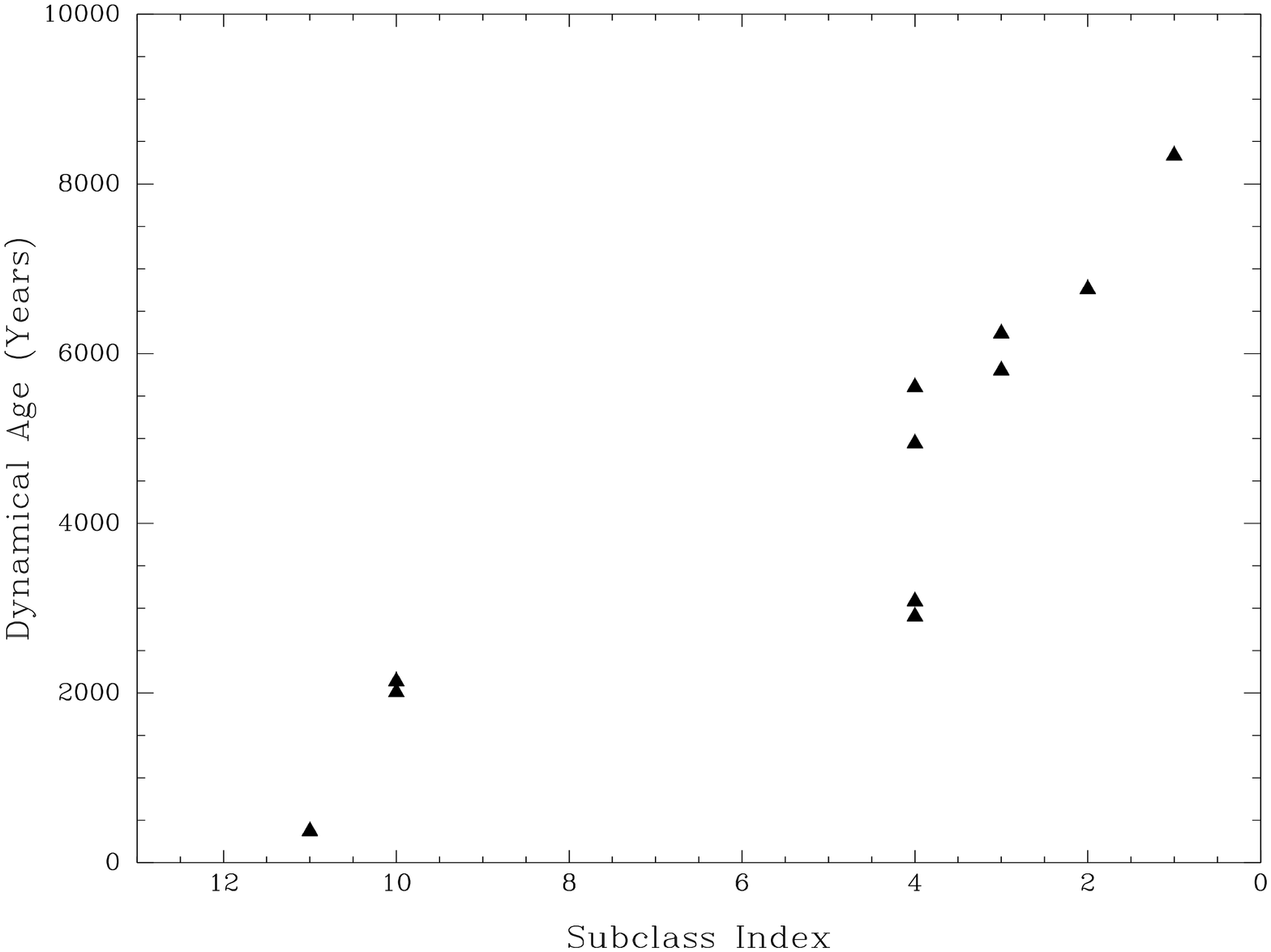}}}
\caption{At left, the spectrum and image of PHR0723+0036, a new MASH PN classifiable as a [WC4].  At right, the preliminary [WR] subclass-dynamical age relationship.  Subclass index corresponds to: [WO1]=1, \ldots, [WO4]=4, [WC4]=5, \ldots, [WC11]=12.}
\end{flushleft}
\label{Figures}
%\end{minipage}
\end{figure}

\section{The [WN] Stars}

Massive WN stars possess significant amounts of hydrogen that will be lost with the surface nitrogen as they proceed toward the WC and WO phases.  The existence of an actual [WN] sequence is open to debate, however.  While hot bottom burning (HBB) should strongly enhance nitrogen in the more massive CSPN progenitors \citep[$\gtrsim$ 4 M$_{\odot}$;][]{2009MNRAS.396.1046L}, there is confusion over whether the two currently designated [WN]s, PM5 for example \citep{2003MNRAS.346..719M} and LMC-N66 \citep{1995RMxAC...3..215P} are truly CSPN.  The high expansion velocity of the main shell of PM5 ($\sim$165 km s$^{-1}$) is far higher than the $v_{exp}$ of other known PN, and is more consistent with a massive WR ring nebula.  LMC-N66 could be a peculiar binary system \citep{2003A&A...409..969H}.

There exists only one published member of the [WN/WC] class, PB 8 \citep{2010A&A...515A..83T}.  The [WN/WC] class exhibits compositions and spectra similar to a massive WR transition type denoted as WN/WC.  These objects have strong nitrogen lines, but also possess significant carbon abundances (1.3\% by mass, as compared to 2\% of nitrogen). Strangely, PB 8 is not a Type I nebula \citep{2010A&A...515A..83T}, as would normally be expected around a high-mass central star which has undergone HBB.  This may not be a completely anomalous result however, as Abell 48 \citep[also not a Type I PN;][]{a48} appears to possess a [WN/WC] CS as well \citep{a48}. 

\section{Towards An Evolutionary Subclass Sequence}

Distance is one of the most crucial and elusive parameters required for a determination of luminosity, mass and other important properties of a star.  The new surface brightness-radius relationship found in \citet{2006IAUS..234...49F} and \citet{frew08} provides a new, robust method for determining distance to PN.  This relation is very easy to use, requiring only the H$\alpha$ flux, the extinction, and the angular dimensions of the PN.

Between 19-23 April 2010, we observed a selection of PN on the Wide Field Spectrograph \citep[WiFeS;][]{2010Ap&SS.327..245D} for chemical abundance determinations at Siding Spring Observatory.  From these observations we obtained the global R$_{[N II]}$ ratio ($\frac{I(\lambda\lambda6548,84)}{I(\lambda6563)}$).  We used this value to deconvolve the [N II] contribution and extract the H$\alpha$ flux using data from the SuperCOSMOS H$\alpha$ Survey \citep[SHS;][]{2005MNRAS.362..689P} and the Southern H$\alpha$ Sky Survey Atlas \citep[SHASSA;][]{2001PASP..113.1326G}.  For those PN with nebular $v_{exp}$ in the literature, H$\alpha$ surface brightness and distance were calculated and a dynamical age derived.  A preliminary plot of [WR] subclass versus dynamical age is shown in Figure 1, where a clear trend is evident.  More in-depth results will follow \citep{depewwels}.  While the analysis is not yet complete, it appears as though the late types are significantly younger than the early types, as previously suspected \citep[e.g.][]{2000A&A...362.1008G}.  However, more data points are needed to firm up this result. 

%\begin{figure*}
%\setlength{\intextsep}{1.0cm}
%\begin{minipage}[t]{13cm}
%\begin{center}
%\includegraphics[clip=true,scale=0.44,angle=270]{dynagecrop.eps} 
%\end{center}
%\end{minipage}
%\caption{\setlength{\baselineskip}{10pt}Preliminary subclass-dynamical age relationship.  The subclass index corresponds to: [WO1]=1, \ldots, [WO4]=4, [WC4]=5, \ldots, [WC11]=12.}
%%\begin{flushleft}
%%\end{flushleft}
%\label{DynAge}
%\end{figure*}

\section{Conclusions}

The provenance of [WR] CSPN is still uncertain.  However, a larger sample size, arising largely from our new discoveries over the past decade, will enable us to fill in some of the gaps in our understanding.  The [WN] and [WN/WC] classes in particular will benefit from new survey data.

In addition, using the SB-r relation of \citet{frew08}, we can determine more accurate distances to PN in our sample, allowing us to establish dynamical ages.  The resulting evolutionary sequence, though somewhat crude, may in the future serve to inform stellar models for this rare class of objects.

\acknowledgements KD thanks Macquarie University for an MQRES PhD scholarship and the Faculty of Science for travel funding for APN V conference attendance.

\bibliography{depew}

\end{document}